\documentclass[a4paper,11pt]{article}
\usepackage{pos}
\usepackage{float}
\usepackage{aps_macros}
\usepackage{listings}
\usepackage{lineno}
\usepackage{array}
\newcolumntype{P}[1]{>{\centering\arraybackslash}p{#1}}
\newcolumntype{M}[1]{>{\centering\arraybackslash}m{#1}}

\title{VTSCat - The VERITAS Catalog of Gamma Ray Observations}
\ShortTitle{VTSCat: VERITAS Data Catalog}

\author*[a]{Sameer Patel}
\author[b]{Gernot Maier}
\author[a]{Philip Kaaret}
\author[+]{VERITAS Collaboration}

\affiliation[a]{University of Iowa,\\
  Department of Physics and Astronomy, Iowa City, USA}

\affiliation[b]{DESY, Platanenallee 6, 15738 Zeuthen, Germany}

\emailAdd{sameer-patel-1@uiowa.edu}
\emailAdd{gernot.maier@desy.de}

\abstract{We present a catalog of results of gamma-ray observations made by VERITAS, published from 2008 to 2020. VERITAS is a ground based imaging atmospheric Cherenkov telescope observatory located at the Fred Lawrence Whipple Observatory (FLWO) in southern Arizona, sensitive to gamma-ray photons with energies in the range of $\sim$ 100 GeV - 30 TeV. Its observation targets include galactic sources such as binary star systems, pulsar wind nebulae, and supernova remnants, extragalactic sources like active galactic nuclei, star forming galaxies, and gamma-ray bursts, and some unidentified objects. The catalog includes in digital form all of the high-level science results published in 112 papers using VERITAS data and currently contains data on 57 sources.  The catalog has been made accessible via GitHub and at NASA's HEASARC.}

\FullConference{37$^{\rm{th}}$ International Cosmic Ray Conference (ICRC 2021)\\
		July 12th -- 23rd, 2021\\
		Online -- Berlin, Germany}

\begin{document}
\maketitle

\section{The VERITAS Observatory}
Located at the Fred Lawrence Whipple Observatory (FLWO) in southern Arizona, VERITAS commenced operations using all four telescopes in 2007. Employing the imaging atmospheric Cherenkov observation technique, VERITAS detects VHE gamma rays through the Cherenkov radiation generated in air showers initiated by the interaction of the primary gamma-rays in Earth's atmosphere. The Cherenkov photons are detected using photomultiplier tubes. The stereoscopic use of 4 telescopes leads to improved shower reconstruction and background rejection.

\begin{figure}
    \includegraphics[width=.99\textwidth]{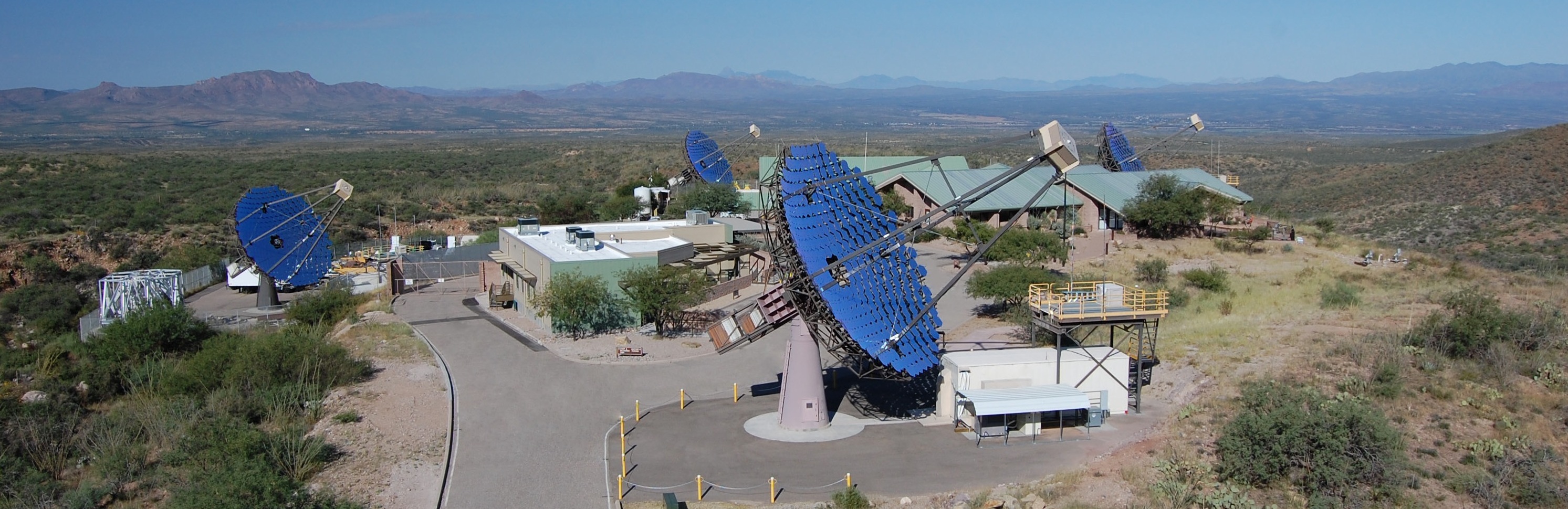}
    \caption{The VERITAS Array at FLWO Basecamp (\href{https://veritas.sao.arizona.edu/}{https://veritas.sao.arizona.edu})}
    \label{fig:vts_array}
\end{figure}

\begin{table}[H]
\centering
\resizebox{\textwidth}{!}{%
% \begin{tabular}{|c|c|c|c|c|}
\begin{tabular}{|m{0.3\linewidth}|c|m{0.2\linewidth}|m{0.3\linewidth}|c|}
\hline
\textbf{Data Type} & \textbf{File Formats} & \textbf{Keyword} & \textbf{Data Description} & \textbf{Reference Figure}                                     \\ \hline
Observation Details   & yaml                  & -                & Significance, spectral models, etc. & Listing \ref{501_obs_data}\\ \hline
Light Curve        & ecsv, png             & lc               & Integrated gamma ray fluxes vs. time & Listing \ref{501_lc_data}; fig. \ref{fig:501_plots}                         \\ \hline
Spectral Energy Distribution & ecsv, png & sed & Spectral flux points with errors and/or upper limits & Listing \ref{501_sed_data}; fig. \ref{fig:501_plots} \\ \hline
Sky Map            & fits                  & signif-skymap, excess-skymap    & Statistical significance or excess sky map    & -                            \\ \hline
\end{tabular}%
}
\caption{Data Formats in VTSCat}
\label{table:data_table}
\end{table}

% \begin{figure}[htp]
% \centering
% \includegraphics[width=.3\textwidth]{figures/501_observation_data.png}
% \includegraphics[width=.3\textwidth]{figures/501_sed_data.png}
% \includegraphics[width=.3\textwidth]{figures/501_lc_data.png}

% \caption{Left: Observation data of Mrk 501 in \cite{mrk_501_plots}. Center: Mrk 501 spectral energy distribution (SED) data (high state) in \cite{mrk_501_plots}. Right: Mrk 501 lightcurve data in \cite{mrk_501_plots}}
% \label{fig:501_data}
% \end{figure}

\begin{lstlisting}[breaklines=true,frame=single,caption=Observation details of Mrk 501 in \cite{mrk_501_plots},label=501_obs_data, basicstyle=\normalsize]
---
source_id: 91
reference_id: 2017A&A...603A..31A
file_id: 1
telescope: veritas

# VERITAS high (TeV flare  2009 May 1)
spec:
  erange: {min: 0.2, max: 6, unit: TeV}
  mjd: {max: 54955., min: 54952.41}
  model:
    type: pl
    parameters:
      norm: {val: 4.17, err: 0.24, scale: 1e-11, unit: cm-2 s-1 TeV-1}
      index: {val: 2.26, err: 0.06}
      e_ref: {val: 1, unit: TeV}
\end{lstlisting}

\section{VTSCat - The VERITAS Data Catalog}
\label{cat}

The VERITAS Data Catalog (hereafter refered to as VTSCat), is a collection of all the data used in more than 100 papers published by the VERITAS collaboration.
The majority of the catalog is compiled from the results of known gamma-ray sources but also includes some results from dark matter studies, measurements of the extragalactic background light, cosmic ray spectra, and upper limits on candidate gamma-ray sources.

\begin{lstlisting}[frame=single,caption=Mrk 501 light curve data in \cite{mrk_501_plots},label=501_lc_data, basicstyle=\normalsize]
# %ECSV 0.9
# ---
# datatype:
# - {name: e_min, unit: TeV, datatype: float64}
# - {name: time, unit: MJD, datatype: float64}
# - {name: flux, unit: cm-2 s-1, datatype: float64}
# - {name: flux_err, unit: cm-2 s-1, datatype: float64}
# meta: !!omap
# - data_type: lc
# - source_id: 91
# - reference_id: 2017A&A...603A..31A
# - telescope: veritas
# - {SED_TYPE: flux}
# - comments: |
#      VERITAS light curve
e_min time flux flux_err
0.30   54907.9710185 6.74707909798e-11 8.84958247235e-12
0.30   54914.8976852 2.97166550106e-11 3.38276780387e-12
0.30   54946.8948611 3.85096002476e-11 9.14725837687e-12
0.30   54951.893044 2.16908417652e-11 5.77378422019e-12
0.30   54952.9114699 1.89636725809e-10 1.22037567454e-11
0.30   54953.7946065 9.6125590747e-11 9.56236260066e-12
0.30   54954.7541319 1.1551067977e-10 2.26421166418e-11
0.30   54955.8014468 8.44464848202e-11 1.85370201643e-11
0.30   54964.6290046 5.97134728089e-11 1.79060782762e-11
0.30   54980.7855324 2.40990545415e-11 8.1438236656e-12
0.30   54982.8297106 2.35129258185e-11 7.3866746951e-12
0.30   54983.8427778 5.14595146513e-11 1.05960376982e-11
0.30   54994.235669 6.64172364739e-11 1.77361601101e-11
0.30   54995.2250903 4.59656118181e-11 1.07883414089e-11
0.30   54996.2994653 4.49888567164e-11 1.02904552945e-11
0.30   54997.2639329 3.12936163865e-11 1.068283848e-11
0.30   54998.2885972 1.99661168475e-11 9.04910537218e-12
0.30   54999.2805185 1.13290032915e-11 8.65266918864e-12
0.30   55001.2647546 1.61178361422e-11 9.25041021514e-12
0.30   55004.1765255 2.75049516655e-11 8.72583430906e-12
\end{lstlisting}

\begin{figure}[!ht]
    \includegraphics[width=.49\textwidth]{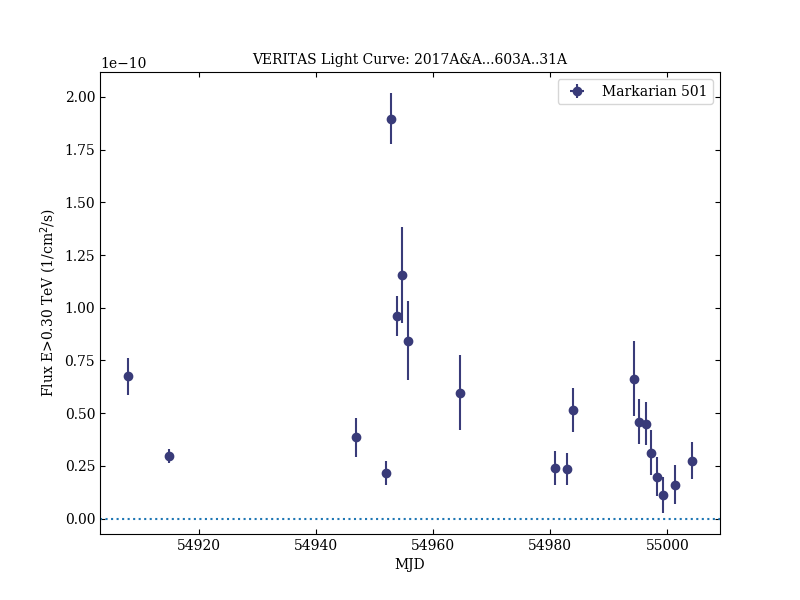}
    \includegraphics[width=.49\textwidth]{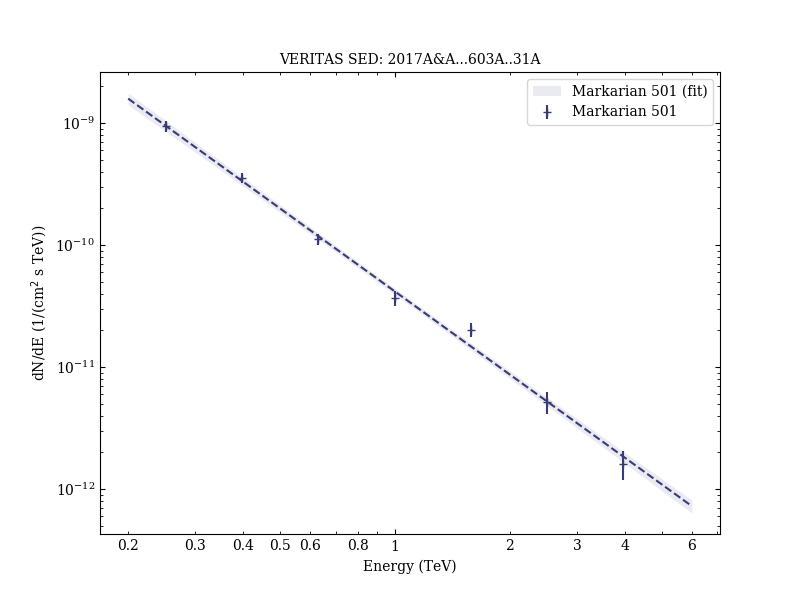}
    \caption{Left: Mrk 501 light curve \& Right: Mrk 501 spectral energy distribution generated from VTSCat data files shown in listings (\ref{501_lc_data} \& \ref{501_sed_data})}
    \label{fig:501_plots}
\end{figure}

\begin{figure}[!ht]
\centering
   \includegraphics[scale=0.35]{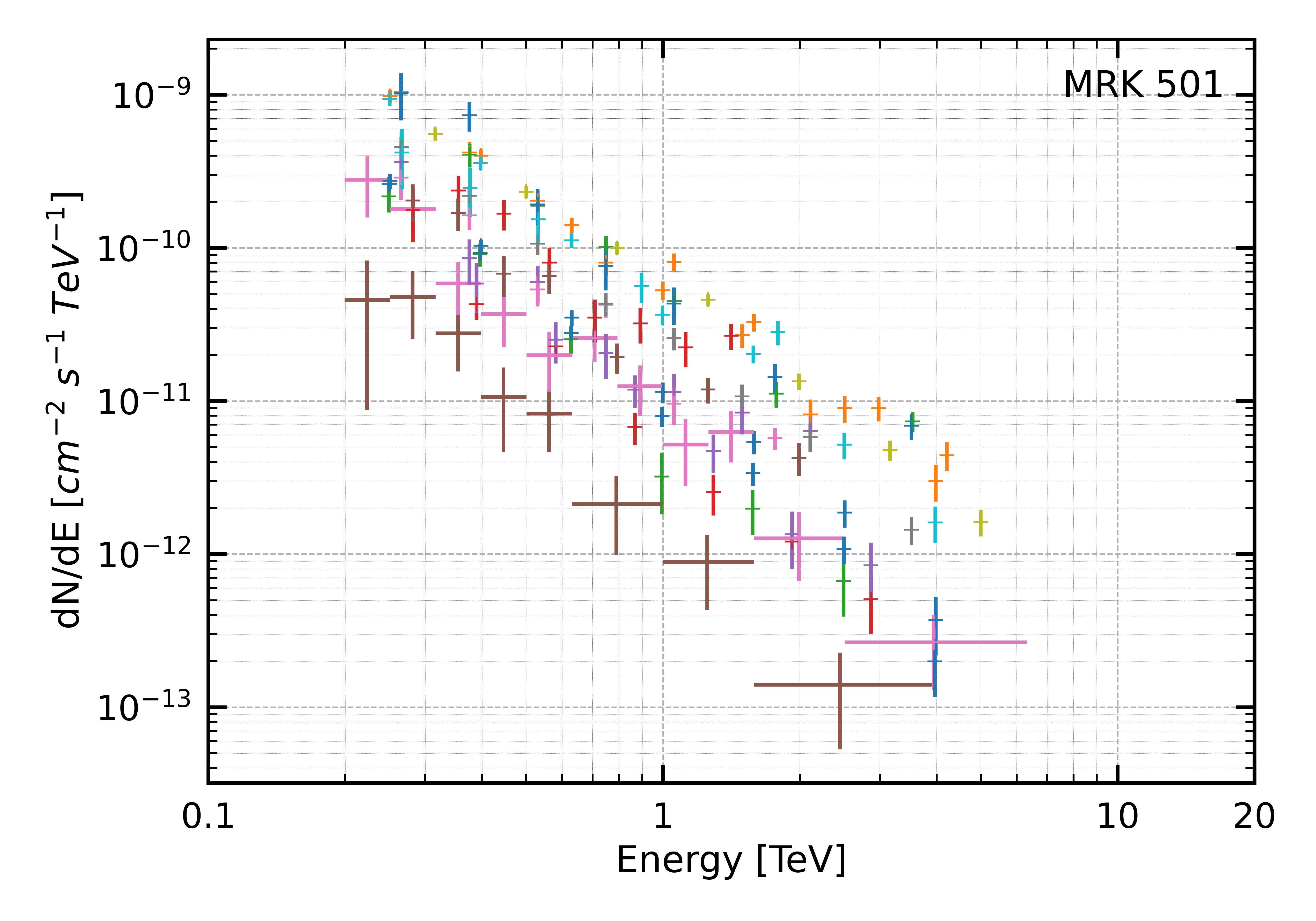}
    \caption{Combined spectral energy distribution (SED) data of Mrk 501 \cite{mrk_501_plots,mrk_501_1,mrk_501_2,mrk_501_3,mrk_501_4, mrk_501_5,mrk_501_6}}
    \label{fig:501_sed}
\end{figure}

Following closely in the footsteps of the structure of GammaCat,\footnote{\href{https://gamma-cat.readthedocs.io/}{https://gamma-cat.readthedocs.io/}} VTSCat consists of high level machine and human readable data files in different formats as described in Table~\ref{table:data_table}. Listings \ref{501_obs_data}, \ref{501_lc_data} \& \ref{501_sed_data} show the formats of some of the data files while Figure~\ref{fig:501_plots} shows quicklook plots. An example use of the VTSCat data is shown in Figure~\ref{fig:501_sed} which presents all of the spectral energy distributions of Mrk~501 measured by VERITAS.

\begin{lstlisting}[frame=single,caption=Mrk 501 spectral energy distribution (SED) data (high state) in \cite{mrk_501_plots},label=501_sed_data, basicstyle=\small]
# %ECSV 0.9
# ---
# datatype:
# - {name: e_ref, datatype: float32, unit: TeV}
# - {name: dnde, datatype: float32, unit: m-2 s-1 TeV-1}
# - {name: dnde_err, datatype: float32, unit: m-2 s-1 TeV-1}
# meta: !!omap
# - data_type: sed
# - source_id: 91
# - reference_id: 2017A&A...603A..31A
# - file_id: 1
# - telescope: veritas
# - comments: |
#      VERITAS high state
e_ref dnde dnde_err
0.2506  9.39e-06 9.12e-07
0.3972  3.57e-06 3.45e-07 
0.6295  1.12e-06 1.23e-07 
0.9977  3.66e-07 5.16e-08 
1.581  2.03e-07 2.69e-08  
2.506  5.18e-08 1.02e-08  
3.972  1.61e-08 4.31e-09
\end{lstlisting}

\clearpage

The data catalog is available for public access at the following data repositories:
\begin{itemize}
    \item GitHub: \url{https://github.com/VERITAS-Observatory/VERITAS-VTSCat/releases}
    \item Zenodo \cite{vtscat}: \url{https://zenodo.org/record/4964083},
    \item HEASARC: to be published under \url{https://heasarc.gsfc.nasa.gov/W3Browse/all/verimaster.html}.
\end{itemize}

Each data file contains the appropriate data units that can be read through Astropy \cite{astropy_1, astropy_2}. The data in VTSCat on Github \& Zenodo has been organized by year and publication, using ADS bibcodes\footnote{\href{https://ui.adsabs.harvard.edu/help/actions/bibcode}{https://ui.adsabs.harvard.edu/help/actions/bibcode}} as reference identifiers. The different objects are then identified by the \texttt{src\_id} (source id), which is an integer value. The description files for these objects can be found in the \texttt{Source} folder. One can then choose the required type of data in the folder by identifying the data file format as in table \ref{table:data_table}. As an example, to get to the lightcurve data of MRK 501 (as in listing \ref{501_lc_data}), the absolute path would be \texttt{/2017/2017A\&A...603A..31A/VER-000091-lc.ecsv}.

\begin{figure}[!ht]
    \includegraphics[width=\textwidth]{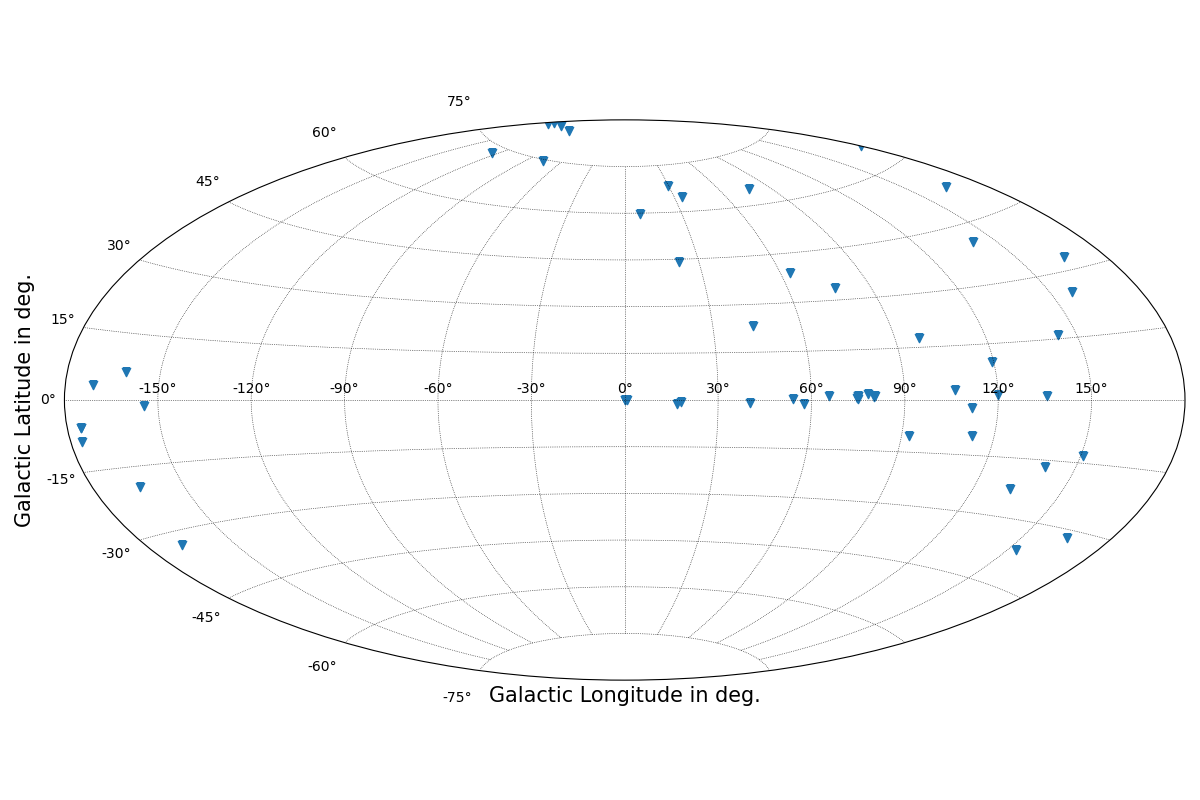}
    \caption[An Aitoff projection of gamma-ray sources detected by VERITAS]{An Aitoff projection of gamma-ray sources detected\footnotemark \ by VERITAS}
    \label{fig:vts_sources}
\end{figure}

\footnotetext{as published in papers}

The \href{https://github.com/VERITAS-Observatory/VERITAS-VTSCat/blob/main/README.md}{README} file provides the catalog observation schema, object names, data files, ADS references, etc. and the \href{https://github.com/VERITAS-Observatory/VERITAS-VTSCat/blob/main/Formats_and_Models.md}{Formats and Models} and \href{https://github.com/VERITAS-Observatory/VERITAS-VTSCat/blob/main/SpectralModels.md}{Spectral Models} files provide information about data types and the spectral models of the yaml files respectively.

\section{Catalog Statistics}

\begin{figure}[!ht]
    \includegraphics[width=.48\textwidth]{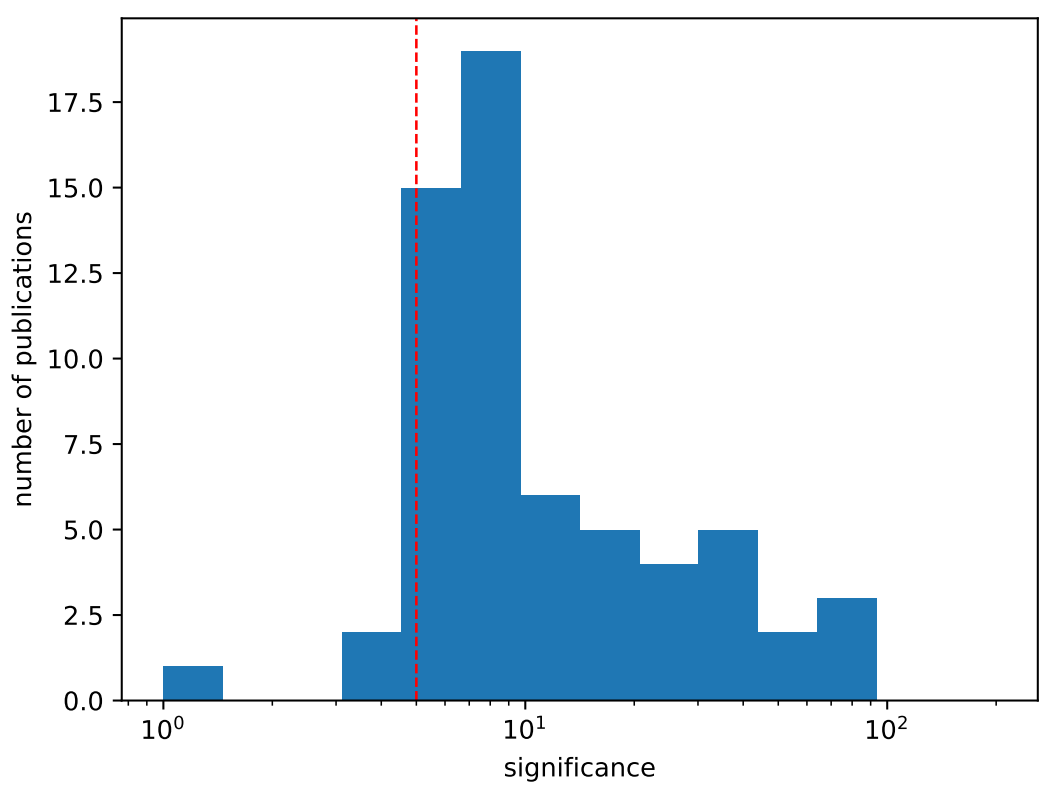}
    \includegraphics[width=.48\textwidth]{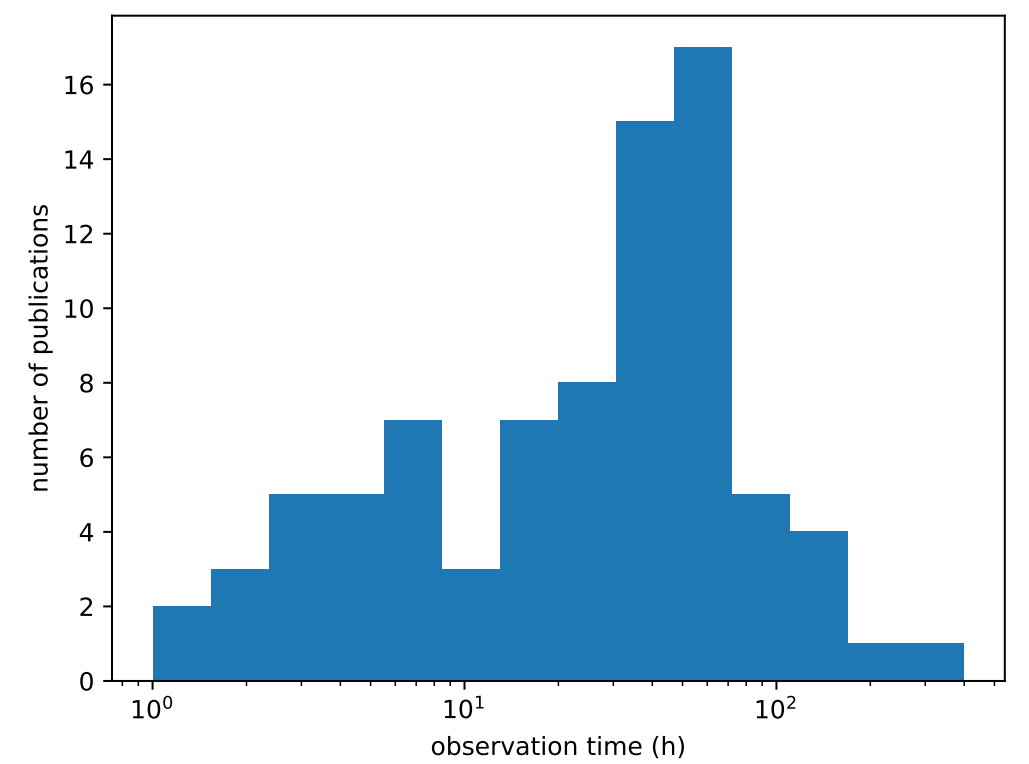}
    \caption{Number of VERITAS publications vs. significance with the dotted line as the VERITAS publication threshold (left) and observation time (right)}
    \label{fig:pub_sig_obs}
\end{figure}

Currently, VTSCat contains data on 57 gamma-ray sources collected between 2008 and 2020, some with multiple observational datasets. The skymap of VERITAS detected sources is shown in Fig.~\ref{fig:vts_sources}. We also include a number of statistical plots relating to published papers in \texttt{ADSdata}, some of which can be seen in Fig.~\ref{fig:pub_sig_obs}. The plot on the left shows the number of publications vs. the statistical significance of the excess of the source (expressed in sigma). Sources with $\sigma <$ 5 include variables sources, sources with upper limit measurements or those without a VERITAS detection of the known TeV sources. The plot on the right shows the number of publications against the observation time of the source (in hours) as reported by authors in the corresponding papers. The catalog repositories listed in section \ref{cat} will be updated periodically as VERITAS collaboration authors publish papers on new and existing VERITAS sources.

\section{Summary}

In this proceeding we have presented the VTSCat, an actively updated, publicly accesible and a machine/human readable catalog of the results from the VERITAS array. It is our hope that in making 12 years of VERITAS data more easily accessible, it can be easily integrated and help to enhance the work of future publications within the community. 

\acknowledgments

This research is supported by grants from the U.S. Department of Energy Office of Science, the U.S. National Science Foundation and the Smithsonian Institution, by NSERC in Canada, and by the Helmholtz Association in Germany. This research used resources provided by the Open Science Grid, which is supported by the National Science Foundation and the U.S. Department of Energy's Office of Science, and resources of the National Energy Research Scientific Computing Center (NERSC), a U.S. Department of Energy Office of Science User Facility operated under Contract No. DE-AC02-05CH11231. We acknowledge the excellent work of the technical support staff at the Fred Lawrence Whipple Observatory and at the collaborating institutions in the construction and operation of the instrument.

\bibliographystyle{jhep}
\bibliography{Bibliography}

\providecommand{\href}[2]{#2}\begingroup\raggedright\begin{thebibliography}{10}

\bibitem{mrk_501_plots}
M.L.~{Ahnen}, S.~{Ansoldi}, L.A.~{Antonelli}, P.~{Antoranz}, A.~{Babic},
  B.~{Banerjee} et~al., \emph{{Multiband variability studies and novel
  broadband SED modeling of Mrk 501 in 2009}},
  \href{https://doi.org/10.1051/0004-6361/201629540}{\emph{\aap} {\bfseries
  603} (2017) A31} [\href{https://arxiv.org/abs/1612.09472}{{\ttfamily
  1612.09472}}].

\bibitem{mrk_501_1}
A.A.~{Abdo}, M.~{Ackermann}, M.~{Ajello}, A.~{Allafort}, L.~{Baldini},
  J.~{Ballet} et~al., \emph{{Insights into the High-energy
  {\ensuremath{\gamma}}-ray Emission of Markarian 501 from Extensive
  Multifrequency Observations in the Fermi Era}},
  \href{https://doi.org/10.1088/0004-637X/727/2/129}{\emph{\apj} {\bfseries
  727} (2011) 129} [\href{https://arxiv.org/abs/1011.5260}{{\ttfamily
  1011.5260}}].

\bibitem{mrk_501_2}
V.A.~{Acciari}, T.~{Arlen}, T.~{Aune}, M.~{Beilicke}, W.~{Benbow},
  M.~{B{\"o}ttcher} et~al., \emph{{Spectral Energy Distribution of Markarian
  501: Quiescent State Versus Extreme Outburst}},
  \href{https://doi.org/10.1088/0004-637X/729/1/2}{\emph{\apj} {\bfseries 729}
  (2011) 2} [\href{https://arxiv.org/abs/1012.2200}{{\ttfamily 1012.2200}}].

\bibitem{mrk_501_3}
J.~{Aleksi{\'c}}, S.~{Ansoldi}, L.A.~{Antonelli}, P.~{Antoranz}, A.~{Babic},
  P.~{Bangale} et~al., \emph{{Multiwavelength observations of Mrk 501 in
  2008}}, \href{https://doi.org/10.1051/0004-6361/201322906}{\emph{\aap}
  {\bfseries 573} (2015) A50}
  [\href{https://arxiv.org/abs/1410.6391}{{\ttfamily 1410.6391}}].

\bibitem{mrk_501_4}
E.~{Aliu}, S.~{Archambault}, A.~{Archer}, T.~{Arlen}, T.~{Aune}, A.~{Barnacka}
  et~al., \emph{{Very high energy outburst of Markarian 501 in May 2009}},
  \href{https://doi.org/10.1051/0004-6361/201628744}{\emph{\aap} {\bfseries
  594} (2016) A76} [\href{https://arxiv.org/abs/1608.01569}{{\ttfamily
  1608.01569}}].

\bibitem{mrk_501_5}
M.L.~{Ahnen}, S.~{Ansoldi}, L.A.~{Antonelli}, P.~{Antoranz}, A.~{Babic},
  B.~{Banerjee} et~al., \emph{{Multiband variability studies and novel
  broadband SED modeling of Mrk 501 in 2009}},
  \href{https://doi.org/10.1051/0004-6361/201629540}{\emph{\aap} {\bfseries
  603} (2017) A31} [\href{https://arxiv.org/abs/1612.09472}{{\ttfamily
  1612.09472}}].

\bibitem{mrk_501_6}
A.~{Furniss}, K.~{Noda}, S.~{Boggs}, J.~{Chiang}, F.~{Christensen}, W.~{Craig}
  et~al., \emph{{First NuSTAR Observations of Mrk 501 within a Radio to TeV
  Multi-Instrument Campaign}},
  \href{https://doi.org/10.1088/0004-637X/812/1/65}{\emph{\apj} {\bfseries 812}
  (2015) 65} [\href{https://arxiv.org/abs/1509.04936}{{\ttfamily 1509.04936}}].

\bibitem{vtscat}
W.~Benbow, A.~Brill, M.~Capasso, J.~Christiansen, M.~Errando, A.~Falcone
  et~al., \emph{{VTSCat: The VERITAS Catalog of Gamma-Ray Observations}},
  Apr., 2021.
\newblock 10.5281/zenodo.4964083.

\bibitem{astropy_1}
{Astropy Collaboration}, T.P.~{Robitaille}, E.J.~{Tollerud}, P.~{Greenfield},
  M.~{Droettboom}, E.~{Bray} et~al., \emph{{Astropy: A community Python package
  for astronomy}},
  \href{https://doi.org/10.1051/0004-6361/201322068}{\emph{\aap} {\bfseries
  558} (2013) A33} [\href{https://arxiv.org/abs/1307.6212}{{\ttfamily
  1307.6212}}].

\bibitem{astropy_2}
{Astropy Collaboration}, A.M.~{Price-Whelan}, B.M.~{Sip{\H{o}}cz},
  H.M.~{G{\"u}nther}, P.L.~{Lim}, S.M.~{Crawford} et~al., \emph{{The Astropy
  Project: Building an Open-science Project and Status of the v2.0 Core
  Package}}, \href{https://doi.org/10.3847/1538-3881/aabc4f}{\emph{\aj}
  {\bfseries 156} (2018) 123}
  [\href{https://arxiv.org/abs/1801.02634}{{\ttfamily 1801.02634}}].

\end{thebibliography}\endgroup

\end{document}